\documentclass[aps,notitlepage,onecolumn]{revtex4-1}
\usepackage{lipsum}
\usepackage{graphicx}
\usepackage{subfigure}
\usepackage{color}
\usepackage{amsmath}
\usepackage[linktocpage,colorlinks=true,linkcolor=blue,citecolor=blue,urlcolor=blue,breaklinks=true]{hyperref}
\usepackage{verbatim}
\usepackage{breakcites}
\usepackage{wrapfig}
\usepackage{soul}
\usepackage{bbold}
\usepackage{gensymb}
\usepackage{textcomp}
\usepackage{amssymb}
\usepackage{mathtools}
\usepackage{xcolor}
\usepackage{verbatimbox}
\usepackage{multibib}

\newcommand{\Lp}{L_+}
\newcommand{\Lm}{L_-}
\newcommand{\Lpm}{L_{\pm}}
\newcommand{\Lpw}{L_+^w}
\newcommand{\Lmw}{L_-^w}
\newcommand{\Lb}{L^b}

\begin{document}
\title{Confinement-induced accumulation and spontaneous de-mixing of microscopic active-passive mixtures}

\author{Stephen Williams$^1$, Rapha\"el Jeanneret$^2$, Idan Tuval$^{3,4}$, Marco Polin$^{1,3,4,*}$} 
\affiliation{ $^1$Department of Physics, University of Warwick, Coventry, CV4 7AL, United Kingdom\\ 
$^2$ Laboratoire de Physique de l'Ecole Normale Sup\'erieure, ENS, Universit\'e PSL, CNRS, Sorbonne Universit\'e, Universit\'e de Paris, F-75005 Paris, France\\ 
$^3$Departament de F\'isica, Universitat de les Illes Balears, 07071 Palma de Mallorca, Spain\\
$^4$Instituto Mediterr\'aneo de Estudios Avanzados, IMEDEA, 07190 Esporles, Illes Balears, Spain}	

\email[Correspondence: ]{mpolin@imedea.uib-csic.es}

\begin{abstract} 

Understanding the out-of-equilibrium properties of noisy microscale systems and the extent to which they can be modulated externally, is a crucial scientific and technological challenge. It holds the promise to unlock disruptive new technologies ranging from targeted delivery of chemicals within the body to directed assembly of new materials. Here we focus on how active matter can be harnessed to transport passive microscopic systems in a statistically predictable way. Using a minimal active-passive system of weakly Brownian particles and swimming microalgae, we show that spatial confinement leads to a complex non-monotonic steady-state distribution of colloids, with a pronounced peak at the boundary. The particles' emergent active dynamics is well captured by a space-dependent Poisson process resulting from the space-dependent motion of the algae. Based on our findings, we then realise experimentally the spontaneous de-mixing of the active-passive suspension, opening the way for manipulating colloidal objects via controlled activity fields. 
 \end{abstract}
 
\maketitle 

\subsection*{Introduction} 
Evolution has enabled living systems to achieve an exquisite control of matter at the microscopic level. From the precise positioning of chromosomes along the mitotic spindle \cite{Civelekoglu-Scholey2010} to the many types of embryonic gastrulation \cite{Haas2018} cells harness their internal and external motility to reach a predictable order despite the stochasticity intrinsic to the microscopic realm \cite{Tsimring2014}. 
Understanding how order emerges in these active systems is a fundamental scientific challenge with the potential to bring disruptive technologies for the macroscopic control of microscopic structures.  
Here we address this problem within a minimal active-passive experimental model system. 

From a physical perspective, living systems fall under the general category of active matter \cite{Ramaswamy2010}, characterised by emergent phenomena including flocking  \cite{Vicsek1995, Bricard2013,Giavazzi2018}, active turbulence \cite{Dombrowski2004,Wensink2012,Wu2017,Doostmohammadi2017} and motility-induced phase separation \cite{Cates2015a,Grobas2021}. 
As our understanding of single-species active systems progresses, attention has started to veer towards more complex cases where components with different levels of activity interact. Indeed  this is often the case for biological active matter. Intracellular activity, for example, can be used for spatial organisation of passive intracellular organelles \cite{Almonacid2015,Cadot2015,Lin2016}; and clustering induced by motility differentials helps bacterial swarms expand despite antibiotic exposure \cite{Beer2019}.
In order to study the emergent properties of these complex systems, an important and phenomenologically rich minimal model is one that mixes active and inert agents.  
Active impurities have been used to alter the dynamics of grain boundaries in colloidal crystals  \cite{VanDerMeer2016,Ramananarivo2019} and favour the formation of metastable clusters in semi-dilute suspensions \cite{Kummel2015c}.
Sufficiently large concentrations of both active and passive species often reveal a rich spectrum of phases depending on the interactions between constituents \cite{McCandlish2012,Stenhammar2015,Takatori2015,Wysocki2016,Weber2016a,Ma2017,Jahanshahi2019,Ilker2020,Rodriguez2020}.

Active baths, where individual passive inclusions are dispersed within an active suspension, are particularly appealing. 
Their conceptual simplicity makes them a natural starting point to develop a statistical theory of active transport \cite{Kaiser2014,Razin2017,Pietzonka2019,Knezevic2020} with potential applications to micro-cargo delivery, micro-actuation \cite{Palacci2013,Koumakis2013c,Wang2019} and nutrient transport \cite{Mathijssen2018b,Guzman-Lastra2021}.
In general, passive particles in homogeneous and isotropic active baths display enhanced diffusion due a continuous energy transfer from the active component via direct collisions and hydrodynamic interactions \cite{Wu2000,Leptos2009c,Valeriani2011,Mino2011a,Kurtuldu2011,Mino2013,Jepson2013,Patteson2016,Jeanneret2016}.
Designing larger objects with asymmetric shapes can then turn active diffusion into noisy active translation or rotation \cite{DiLeonardo2010,Sokolov2010,Kaiser2014}. 
However, in contrast to the exquisite level of external control possible on biological and synthetic active suspensions \cite{Palacci2013b,Kirkegaard2016b,Arrieta2017,Frangipane2018,Arlt2018,Lozano2016}, the strategies for the predictable patterning and transport of passive cargo are still very limited. 

Here we show that a steady gradient of activity within active-passive suspensions can be harnessed to control the fate of generic passive particles. 
Quasy-2D microfluidic channels are filled with a binary suspension of polystyrene colloids and unicellular biflagellate microalgae {\it Chlamydomonas reinhardtii} (see Movie~M1). Confinement induces a spatially inhomogeneous and anisotropic distribution of microswimmers as a result of wall scattering \cite{Kantsler2013,Contino2015,Ostapenko2018,Matteo_thesis}, which translates into a space-dependent active noise for the colloids. 
We show that this produces complex non-monotonic colloidal distributions with accumulation at the boundaries and then develop an effective microscopic jump-diffusion model for the colloidal dynamics. The latter shows excellent analytical and numerical agreement with the experimental results. Finally we demonstrate how confinement with aptly designed microfluidic chips can fuel the spontaneous de-mixing of active-passive suspensions, opening the way for manipulating passive colloidal objects via controlled activity fields.

\subsection*{Results}

\begin{figure}
\centering
\includegraphics[width=0.5\linewidth]{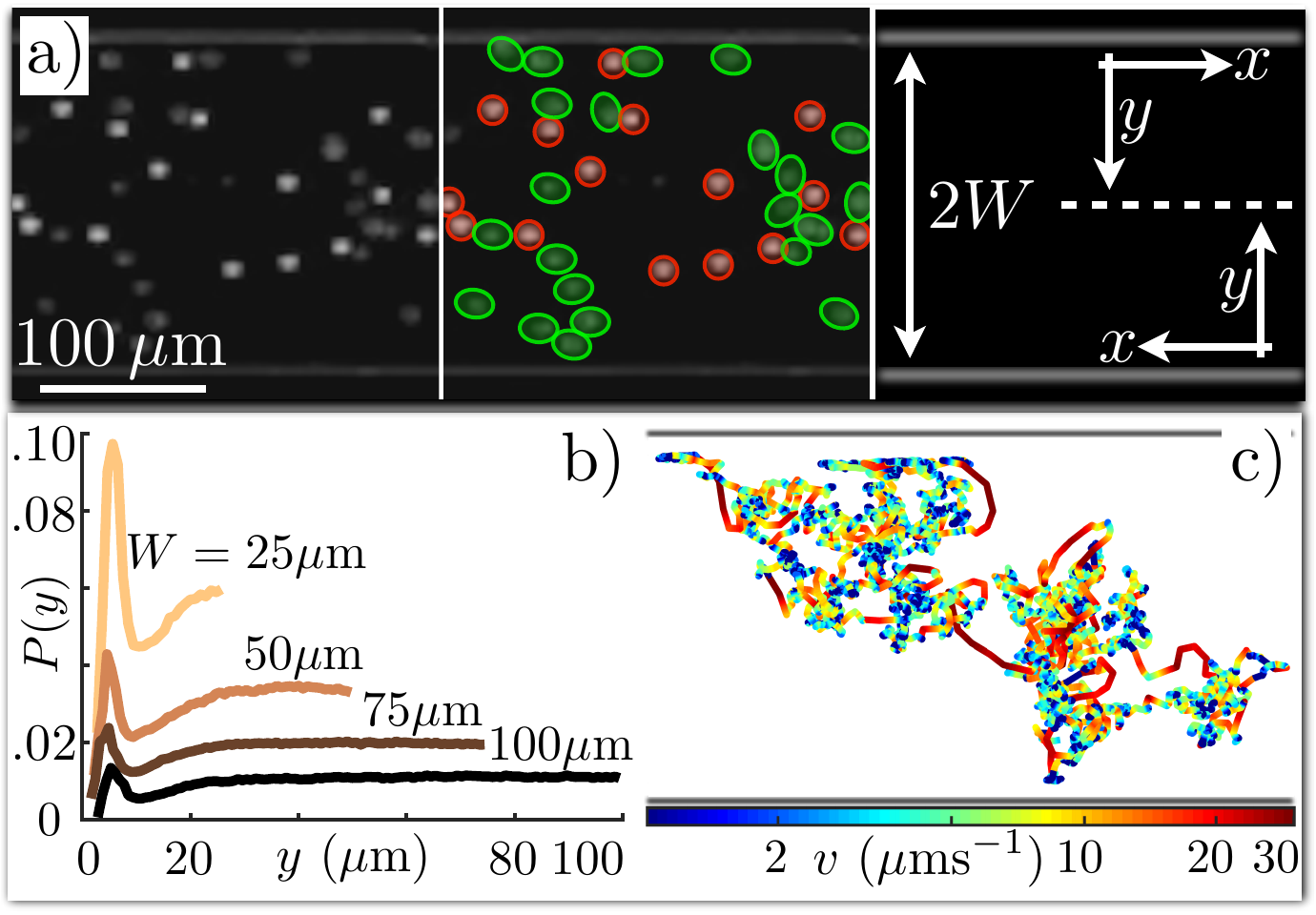}
\caption{\textbf{Experimental setup and colloidal behaviour}. \textbf{a)} Dark-field image of cells and beads within a straight microfluidic channel (here width $2W=200\,\mu$m). The two species are highlighted with, respectively, green ellipses and red circles. The schematic illustrates the coordinate system used throughout. \textbf{b)} Steady state colloidal probability distribution functions across the channel (edge to midpoint) for different $W$ values. Boundary accumulation is followed by a depleted adjacent region which gives way to a uniform distribution within the channel. \textbf{c)} Example of a typical colloidal trajectory inside a $100\,\mu$m-width channel. Colours represent the mean frame-to-frame speed. The dynamics is composed of slow diffusive-like motion (blue sections) interspersed with fast straighter jumps (red sections).}
\label{fig1}
\end{figure}

A full description of the experimental procedures (culturing, video microscopy and data analysis) can be found in the Methods section and Supplementary Materials. Figure~\ref{fig1}a shows a detail of the main section of the first type of microfluidic setup used. These devices, between $14\,\mu$m and $20\,\mu$m thick, are composed of two reservoir chambers connected by long and straight channels of constant width $2W$ ranging from $50$ to $200\,\mu$m.  These are filled with a dilute mixture of {\it Chlamydomonas reinhardtii} (CR; strain CC-125, radius $R\sim4-5\,\mu$m) and weakly Brownian colloids (radius $a=5\pm0.5\,\mu$m) at surface fractions $\phi_{\rm CR}\approx\phi_{\rm col}\approx 2-3\%$ (bulk concentrations $\sim 2-3\times10^7\,$particles/mL). The design ensures a steady concentration along the connecting channels for both species (see Movie M1). 
Figure~\ref{fig1}a shows the coordinate system employed, which has been symmetrized with respect to the channel axis.

As typical of self-propelled particles, the algae tend to accumulate at the boundaries due to their interactions with the side walls \cite{Kantsler2013,Contino2015,Ostapenko2018}. This appears as a significant peak in their time-averaged concentration profile at a position $y_{\rm CR}\approx 15\,\mu$m (Fig.~S1), roughly equivalent to the sum of the cell radius and the flagellar length. 
Unexpectedly, we find that also the colloids explore the available space in an inhomogeneous way. Their steady-state distribution (Fig.~\ref{fig1}b) shows a clear peak at about one particle radius ($y_{\rm col}=5.9 \pm 1.5\,\mu$m), followed by a depleted region between $10\,\mu$m and $20\,\mu$m from the wall, which roughly corresponds to the peak in algal density, before plateauing to a uniform concentration further inside the channels.
This effect is also observed within circular chambers (Fig.~S2) suggesting it is a robust feature of the system.
These distributions are in stark contrast with the equilibrium case (i.e. without micro-swimmers) for which the colloids are expected to be uniformly distributed despite spatial variations in colloidal diffusivity due to hydrodynamics \cite{Lancon2001}.

As a first step to gain insight into the colloids' experimental distributions, we characterise their dynamics within our active bath. 
Figure~\ref{fig1}c shows a typical colloidal trajectory ($561\,$s) colour-coded for the average frame-to-frame speed $v$.
The dynamics can be understood as a combination of periods of slow diffusive-like displacements ($v\sim 5\,\mu$m/s;)
and fast longer jumps ($v\sim 30\,\mu$m/s; see SM Sec.~S1). The latter is reminiscent of hydrodynamic entrainment events reported for micron-sized colloids  \cite{Jeanneret2016, Mathijssen2018}, although for these larger particles the prominent role appears to be played by direct flagellar interactions (see Movie S2).  Irrespective of their specific origin, jump events dominate the active transport of colloids within the system.
\begin{figure}
\centering
\includegraphics[width=0.5\linewidth]{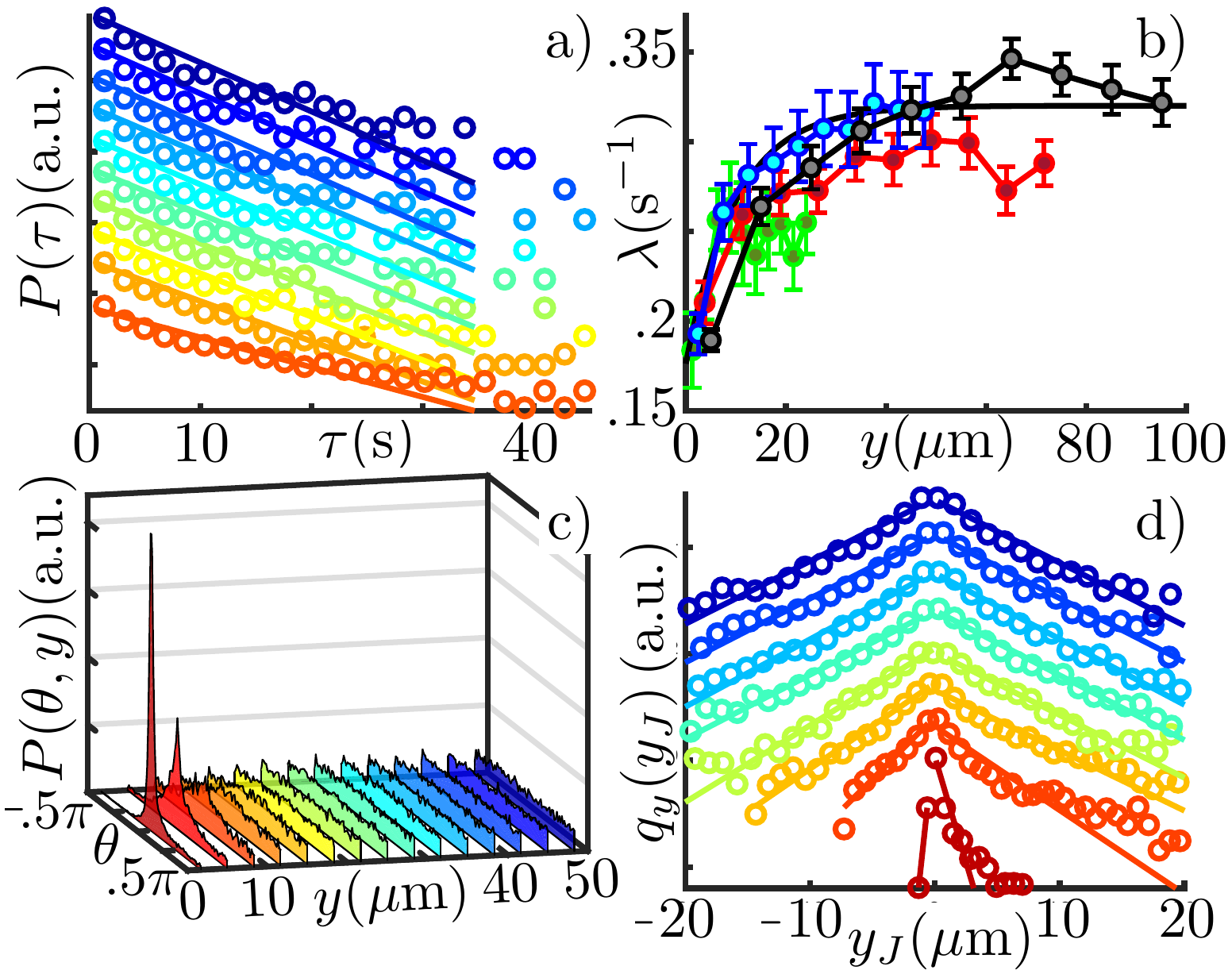}
\caption{\textbf{Colloidal jump dynamics.} \textbf{a)}  Probability distribution function of waiting times between consecutive jumps ($W=50\,\mu$m, semilog plot) ($\bigcirc$) Experiments. ({\bf -}) Exponential fit. The distributions are shifted vertically for clarity. Colours code for position across the channel (red: boundary; blue: channel centre). \textbf{b)} Characteristic encounter rates $\lambda(y)$ for the four values of $W$, as a function of distance from the channel boundaries. For better comparison, the values have been rescaled to the cell concentration of the $100\,\mu$m-wide channel (see SM Sec.~S3). ($\bigcirc$) Experiments. ({\bf -}) Exponential fit (see SM Sec.~S4). \textbf{c)} Probability distribution function $P(\theta,y)$ for the direction $\theta$ of active jumps vs. distance $y$ from the channel boundary ($W=50\,\mu$m). \textbf{d)} Semilog plot of active jump distributions, $q_y(y_{\rm J})$, for different distances $y$ from the boundary (Colours as in C). ($\bigcirc$) Experiments. ({\bf -}) Exponential fits. The distributions are shifted vertically for clarity.}
\label{fig2}
\end{figure}
Following \cite{Jeanneret2016,Mosby2020}, we use time-correlation of displacements along individual trajectories and an estimate for the expected magnitude of frame-to-frame diffusive displacements to identify active colloidal jumps and extract their statistical properties as a function of starting position $y$ (we assume translational invariance along the channels' axis). 
Figure~\ref{fig2}a shows the space-dependent distributions of waiting times between successive jumps, with colours representing the distance from the boundary. Regardless of the position, all the distributions are exponential above $\sim 3\,$s but deviate from it at shorter times. This deviation from a simple Poisson process is due to the large colloidal size influencing the motility of nearby algae as  reported also in the case of bacteria \cite{Lagarde2020}. Here, this can lead to a rapid succession of interactions with the same cell (see Movie~S3). Nevertheless, for our purposes, the waiting dynamics can still be approximated with a single effective rate $\lambda(y)$ (Fig.~\ref{fig2}a solid lines; for further details see SM Sec.~S2).
Figure~~\ref{fig2}b shows that these rates increase monotonically with increasing distance from the wall.
Notice that this curve does not mirror the algal accumulation at the boundary (Fig.~S1), a reflection of the fact that proximity to the wall curtails the range of possible swimming directions that algae can have when interacting with the colloid. 

Next we look at the modulation in jump orientation and magnitude. Figure~\ref{fig2}c shows the distribution of jump directions  $P(\theta,y)$, with motion towards or away from the boundary corresponding to negative and positive values of $\theta$ respectively. Approaching the wall, the distribution develops a marked peak at $\theta=0$ indicating a strongly anisotropic active motion preferentially parallel to the boundary. This feature reflects the anisotropy in algal dynamics that results from the interaction with the wall \cite{Kantsler2013} and that can be measured up to $\sim100\,\mu$m from the boundary because of the persistence in cells' trajectories \cite{Matteo_thesis}. The peak in $P(\theta,y)$ decays exponentially with distance from the wall with a characteristic length $L_{\theta}=16.4\pm0.3\,\mu$m (Fig.~S3).
The distribution functions of jump magnitudes are similar to those already reported for micron-sized particles \cite{Jeanneret2016, Mathijssen2018}, with an exponential decay above $\sim 4\,\mu$m (Fig.~S4a). These can be used to calculate the average jump length $\langle l(y)\rangle$ (Fig.~S4b) which decreases by $\sim 30\%$ from the bulk level within the first $10-15\,\mu$m from the boundary (first $5-10\,\mu$m accessible to the beads).
A decrease is indeed expected due to the obvious limitations to the active movement of the colloids imposed by a nearby boundary.

\begin{figure}
\centering
\includegraphics[width=0.5\linewidth]{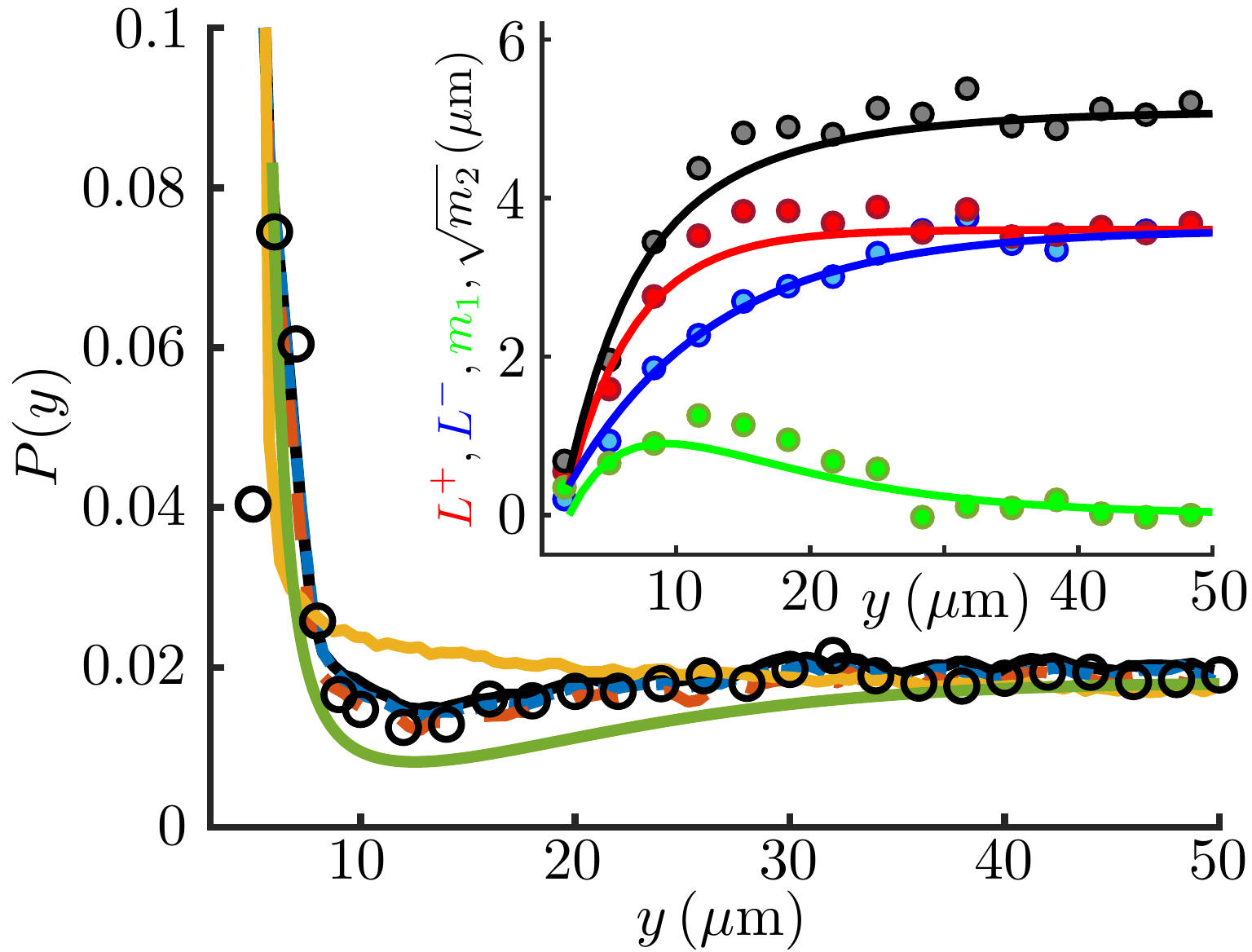}
\caption{\textbf{Comparison between experimental results and the jump-diffusion model}. Steady state colloidal probability distribution functions in a $2W=100\,\mu$m channel: experiments ($\bigcirc$); full dynamics (black solid line {\bf -}); constant encounter rate ($\lambda(y)\coloneqq$bulk value) (teal dashed line {\color{teal}{\bf - -}}); no thermal diffusion ($D_0=0$) (purple dashed line {\color{purple}{\bf - -}}); homogeneous and isotropic jump size distribution ($q_{y}(y_{\rm J})\coloneqq$bulk distribution) (orange solid line {\color{orange}{\bf -}}); analytical model (olive solid line {\color{olive}{\bf -}}). The analytical model curve has been shifted downward for clarity. {\bf Inset}: characteristic jump size away from the boundary ($\Lp(y)$, red circles) and toward the boundary ($\Lm(y)$, blue circles) together with the first and second moments $m_1(y)$ and $\sqrt{m_2(y)}$ derived from them (Eq.~\ref{moments}). Solid lines are fitted heuristic analytical functions used to reconstruct the colloidal distribution (see SM Sec.~S4).}
\label{fig3}
\end{figure}

As we are interested in the passive particles' dynamics across the channels, it is useful to reduce the full two-dimensional jump distribution functions to their projection $q_y(y_{\rm J})$ along the $y$-axis. Here $y_{\rm J}$ is the $y$-coordinate of the active displacement and the subscript indicates the distance from the nearest wall (see Fig.~\ref{fig1}a for the frame of reference).
Figure~\ref{fig2}d shows that these distributions are generally well approximated by the combination of two exponentials, one each for positive and negative directions (respectively away from and towards the boundary). The exception is for the positive tails of the distributions closest to the wall, which decay slower than expected from the exponential fit. They will not be considered in the following. 
The distributions $q_y(y_{\rm J})$ can then be approximated as 
\begin{equation}
\label{distri_jump}
 q_y(y_{\rm J})=\frac{1}{\Lp(y)+\Lm(y)}e^{-|y_{\rm J}|/\Lpm(y)},~0\lessgtr y_{\rm J},
\end{equation}
where $\Lpm(y)$ are the characteristic lengths of the exponential fits to positive and negative jumps respectively (see Sec.~S3 of the SM for details on how the characteristic lengths were extracted). As shown in Fig.~\ref{fig3}-inset (red and blue symbols and solid lines) these are identical in the core of the channel ($\Lp=\Lm\simeq 4\,\mu$m) and decrease to the same small value close to the wall ($\sim 0.5\,\mu$m) as a consequence of the increasing polarisation of the active displacements along the boundary (Fig.\ref{fig2}c). However, the transition between the two values is more abrupt for $\Lp(y)$ than for $\Lm(y)$ ($\sim12\,\mu$m vs. $\sim25\,\mu$m; see SM Sec.~S4). This gives rise to a net drift $\Lp(y)-\Lm(y)>0$ towards the bulk of the system (green symbols and solid line, Fig.~\ref{fig3}-inset) which, as discussed below, is responsible for the depleted region observed in the experimental colloidal distributions.

The focus on the active jumps displayed by the passive particles can only be justified if this part of the dynamics is indeed sufficient to capture the experimental steady state distributions. This was first tested within a 1D numerical simulation of a weakly Brownian particle (diffusivity $D_0$), moving in $y\in[0,2W]$ and subject to a space-dependent Poisson noise of rate $\lambda(y)$ and value drawn from $q_y(y_{\rm J})$ (see Methods Sec.~\ref{methods:simu} for details on the integration scheme). Figure~\ref{fig3} shows that the resulting steady-state distribution (black solid curve) is in excellent agreement with the experimental one (black circles) ($2W=100\,\mu$m). 
Numerical simulations also provide a convenient way to explore which elements of the effective colloidal dynamics play the most prominent role. For example, fixing $\lambda(y)$ to the bulk value everywhere leads to a very minimal change in the spatial distribution (Fig.~\ref{fig3}, teal dashed line), showing that the spatial dependence of the jump frequency close to the wall is not a major factor in the present case. Similarly, removing completely the background diffusion by setting $D_0=0$  leaves the distribution unchanged (Fig. \ref{fig3}, purple dashed line) consistent with the fact that, in our experimental system, colloidal transport is dominated by the Poissonian jumps.
On the other hand, simulations that remove the spatial dependence of the jump distributions $q_{y}(y_{\rm J}$) and use their isotropic bulk value everywhere, show a  boundary accumulation of colloids that is narrower and higher than the experimental curve and has no intermediate depletion (Fig.~\ref{fig3}, orange solid line). This confirms that space-dependent jump anisotropy plays a key role in determining the experimental colloidal distribution. 

The numerical validation of the jump-diffusion dynamics motivates a simple analytical model for the evolution of $P_t(y)$, the probability density of finding a colloid at position $y$ at time $t$:
\begin{align}
\begin{split}
\label{unbounded_FP}
\frac{\partial P_{t}(y)}{\partial t}&=D_0\frac{\partial^2 P_{t}(y)}{\partial y^2}-\lambda(y)P_t(y) \\
& +\int_{-\infty}^{+\infty}\lambda(y-y_{\rm J})P_t(y-y_{\rm J})q_{y-y_{\rm J}}(y_{\rm J})dy_{\rm J}. \\
\end{split}
\end{align}
In this master equation, local changes in $P_t(y)$ are due either to diffusion (first term in the r.h.s) or to the balance between active jumps from the current position or towards it from elsewhere (second and third terms in the r.h.s respectively).
This continuum equation can also be derived more formally from a stochastic description of the dynamics of single colloids, following Denisov and Bystrik \cite{Denisov2019a} (see SM Sec.~S5).
It is worth noting also that Eq.~\ref{unbounded_FP} assumes infinitely fast jumps (``teleportation''). While this might seem like a drastic approximation, it is not expected to impact the resulting spatial distributions as long as the typical jump duration is sufficiently shorter than the average waiting time between jumps. This is indeed the case in the current system ($\sim 1.2\,$s vs. $\gtrsim 3.5\,$s respectively; see SM Fig.~S5). 

%
The non-local term in Eq.~\ref{unbounded_FP} means that the model is non-tractable for generic kernels $\lambda(y)q_{y}(y-y_{\rm J})$. In order to make progress we therefore perform a Kramers-Moyal expansion \cite{Risken1989} and truncate the series at second order.
This approximation is expected to hold if the characteristic jump size is small compared to the length scale of the heterogeneities in the dynamics. In our case we have $\langle l_{\rm bulk}\rangle\approx 5\,\mu$m (Fig.~S4b) while  a length scale for the heterogeneity $L_{\theta}\approx16\,\mu$m can be extracted from the exponential decay of the peak value in the distribution of jump angles (Fig.~S3). 
As detailed in the Supplementary Materials (Sec.~S6) the second-order expansion of Eq.~\ref{unbounded_FP} gives a drift-diffusion equation which depends on  $\lambda(y)$ and the first and second moments $m_{1,2}(y)$ of $q_{y}(y_{\rm J})$:
\begin{align}
\begin{split}
\label{approx_unbounded_FP}
\frac{\partial P_{t}(y)}{\partial t}&=\frac{\partial}{\partial y}\left[D_{\rm eff}(y)\frac{\partial P_t}{\partial y}-V_{\rm eff}(y)P_t(y)\right]\\
D_{\rm eff}(y)&=D_0+\frac{\lambda(y)m_2(y)}{2}\\
V_{\rm eff}(y)&=\lambda(y)m_1(y)-\frac{1}{2}\frac{\partial}{\partial y}\left[\lambda(y)m_2(y)\right]\\
m_n(y)&=\int_{-\infty}^{+\infty}y_{\rm J}^n q_{y}(y_{\rm J})dy_{\rm J}.
\end{split}
\end{align}
Notice that $D_{\rm eff}$ is similar to the asymptotic diffusivity obtained in \cite{Jeanneret2016} for homogeneous and isotropic entrainment-dominated transport, with a contribution from the jump events given by the product of jump frequency and variance of jump size. 
Enforcing no-flux at the boundaries one obtains a closed form for the steady-state solution: 
\begin{align}
\label{gen_sol}
P(y)=\frac{B}{D_{\rm eff}(y)} \exp\left({\int^{y}\frac{\lambda(y')m_1(y')}{D_{\rm eff}(y')}dy'}\right),
\end{align}
where $B$ is the normalisation constant. Without a net asymmetry in the dynamics (i.e. when $m_1(y)\equiv0$), the solution is that of a state-dependent diffusive process with It\^o convention for multiplicative noise integration. In our case, however, the first moment of $q_y$ takes significant values in the first $\sim 30\,\mu$m from the boundaries (green solid line Fig.~\ref{fig3}-Inset) and should not be neglected. 

Comparison between Eq.~\ref{gen_sol} and the experiments is facilitated by having analytical expressions for $\lambda$, $m_1$ and $m_2$. Figure~\ref{fig2}b shows that $\lambda$ can be described heuristically by a simple exponential relaxation from the wall to the bulk values (black solid line).  
As for $m_1$ and $m_2$, the description of $q_y(y_{\rm J})$ given by Eq.~\ref{distri_jump} implies that 
\begin{align}
\label{moments}
\begin{split}
m_1(y)&=\Lp(y)-\Lm(y)\\
m_2(y)&=2\,\frac{\Lp^3(y)+\Lm^3(y)}{\Lp(y)+\Lm(y)}.
\end{split} 
\end{align}
Given that $\Lpm(y)$ themselves are well approximated by exponentials, this affords analytical approximations also for $m_{1,2}(y)$ (Fig.~\ref{fig2}c-Inset).
Armed with this description we can now compare the experimental curves with the prediction from the  approximate jump-diffusion model. 
The olive-colored curve in Fig.~\ref{fig3} shows that the theoretical distributions recapitulate extremely well the experimental ones. On one hand, this provides a justification {\it a posteriori} for the modelling approach taken above. 
On the other it reveals that, at a continuum level, the dynamics of colloids in an active bath can be reduced to two quantities: the first and the second moments of the active displacements. This provides an intuitive way to conceptualise the complex dynamics of colloidal particles within an active suspension.

\begin{figure}
\centering
\includegraphics[width=0.5\linewidth]{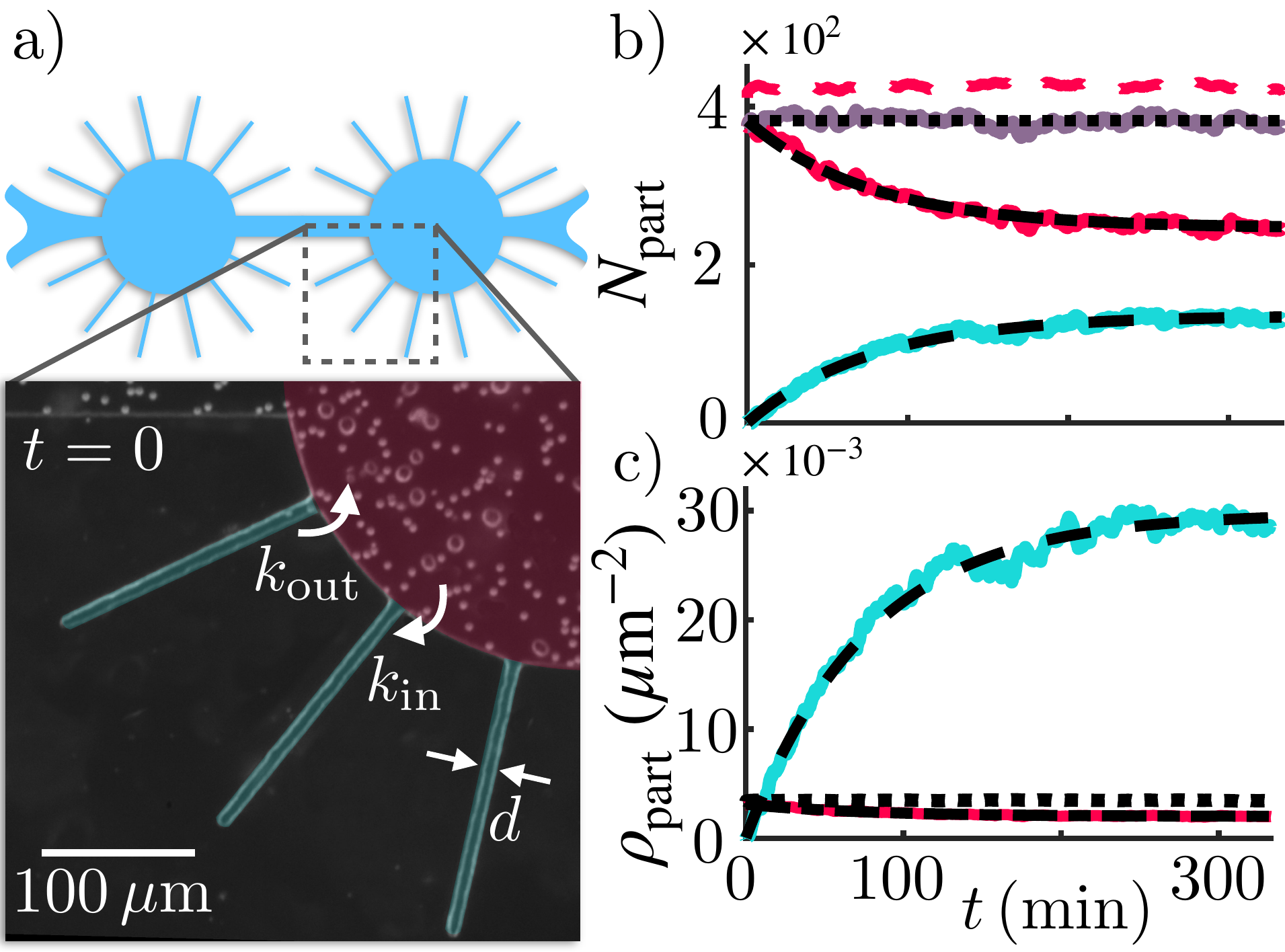}
\caption{\textbf{Spontaneous de-mixing of a binary mixture}. \textbf{a)} Schematics and details of the microfluidic chamber: $200\,\mu$m-radius circular chambers decorated with side channels of width $d=7-8\,\mu$m. Both cells ($2R=8-10\,\mu$m) and polystyrene beads ($2a=6\,\mu$m) are visible in the picture. \textbf{b)} and \textbf{c)} Time evolution of the number of particles and surface number density within the circular chamber ($N_c,\rho_c$ red solid line {\color{red}{\bf -}}) and the side channels ($N_s,\rho_s$ cyan solid line {\color{cyan}{\bf -}}) in presence of the microalgae. Within this time the total number of particles in the chamber remains constant (purple solid line {\color{purple}{\bf -}}). Dashed/dash-dot lines are a fit to the first order kinetics (see SM Sec.~S6). Dotted lines indicate the initial values of $N_c$ and $\rho_c$. Dashed red line  {\color{red}{\bf - -}} corresponds to the control experiment without micro-algae.}
\label{fig4}
\end{figure}

What we learned so far on the colloids' emergent active dynamics can be harnessed to induce the spontaneous de-mixing of our active-passive suspensions, as a first step towards more complex control of microscale cargo transport.
The idea is to include a confining boundary which acts as a kind of selective membrane, letting the beads cross but not the algae. This is achieved by considering circular chambers decorated with $190\,\mu$m-long side channels whose width $d$ (Fig.~\ref{fig4}a) is smaller than the swimmer diameter but larger than the bead diameter (here taken as $2a=6\,\mu$m$\,<d=7-8\,\mu$m$\,<2R=8-10\,\mu$m).  Due to the active dynamics of the colloids, we expect them to be pushed inside the side channels by the algae, therefore depleting the particles in the main chamber and spatially separating the active and passive species.
As shown Fig.~\ref{fig4}b (see also Movie~S4) this is exactly what we obtain, with the number of particles in the circular chamber $N_{\rm c}$ (red solid line) quickly decreasing and the number in the side channels $N_{\rm s}$ (cyan solid line) increasing. Notice that the total number of particles $N_{\rm t}=N_{\rm c}+N_{\rm s}$ in the system (purple solid line) remains constant over the duration of the experiment. 
The process follows a first order kinetic with rates $k_{\rm in}$ and $k_{\rm out}$ for transitions towards or out-of the side channels respectively (see SM Sec.~S7). The occupancy numbers follow an exponential relaxation with time-scale $1/(k_{\rm in}+k_{\rm out})= 77\pm 4\,$min and stationary values $N_{\rm c}^{\rm \infty}/N_{\rm t}=k_{\rm out}/(k_{\rm in}+k_{\rm out})=0.64\pm 0.02$ and $N_{\rm s}^{\rm \infty}/N_{\rm t}=k_{\rm in}/(k_{\rm in}+k_{\rm out})=0.36\pm 0.02$, leading to estimates for the two rates of $k_{\rm in}=(7.8\pm 0.6)\times 10^{-5}\,$s$^{-1}$ and $k_{\rm out}=(13.9\pm 0.8)\times 10^{-5}\,$s$^{-1}$. The latter results from the passive dynamics of the colloids within the side channels, which depends on their thermal diffusivity, potential electrostatic interactions with the boundaries, short-range hydrodynamic interactions between the beads \cite{Misiunas2015}, channel design, etc \cite{Locatelli2016a}. 
The former, instead, is the consequence of active colloidal displacements. It encapsulates the intrinsic out-of-equilibrium nature of the system, a property which is immediately clear from the significant difference in steady-state densities between the side channels ($\rho_{\rm s}=0.0295\pm 0.007\,$beads$/\mu$m$^2$) and in the main chamber ($\rho_{\rm c}=0.0019\pm 0.0001\,$beads$/\mu$m$^2$) (Fig.~\ref{fig4}c). The value of $k_{\rm in}$ can be compared with an estimate derived from the effective colloidal dynamics from Eq.~\ref{approx_unbounded_FP} (see SM Sec.~S8). The mean first passage time of a colloid to the entrance of any of the side channels returns a rate of $k_{\rm in}^{V,D}=(7.408\pm 0.1)\times 10^{-5}\,$s$^{-1}$ which agrees very well with the experimental value. This shows that the simplified advection-diffusion model for the passive particles is a good description not just for time-independent properties like the steady state distribution of Fig.~\ref{fig3}, but also for dynamic ones like the rate of active cargo transport to the side channels. 
It is instructive also to use the model to explore the contributions of the different parts of the dynamics. Hence we see that setting $V_{\rm eff}\equiv0$ and $D_{\rm eff}\equiv 3.55\,\mu$m$^2$s$^{-1}$, the effective diffusivity far from the boundaries, one obtains a rate $k_{\rm in}^{D}=(16.2\pm 4.1)\times 10^{-5}\,$s$^{-1}$.
The stark difference between $k_{\rm in}^{V,D}$ and $k_{\rm in}^{D}$ suggests that, in our case, it is not sufficient to know the behaviour of the active-passive suspension in the bulk of the chamber. The behaviour close to the boundary, where the inhomogeneities are found, is essential to grasp the dynamics of the de-mixing process.
Repeating the same experiment only with colloids shows indeed no appreciable filling of the side channels within the duration of the experiment ($\sim 10\,$h; see Movie~S5) and the particle density in the main chamber remaining constant (Fig.~\ref{fig4}b red dashed line). 

Although the microfluidic chip in Fig.~\ref{fig4} leads to a much higher concentration of colloids in the side channels than in the main chamber, this still involves only $\sim36\%$ of the colloids. A different design, however, could maximise the fraction of de-mixed colloids instead of their concentration. For example, modifying the current design to have two identical chambers connected by a narrow gap of size $d=7-8\,\mu$m should produce a fraction of de-mixed colloids  given by $k_{\rm in}/\left(k_{\rm in}+k_{\rm in}^{\rm th}\right)\simeq 95\%$. The design could then be tailored to different needs.

\subsection*{Discussion}

In this study we used an active-passive system as a minimal model to investigate the active transport of passive microscopic objects.
Leveraging the well-known boundary accumulation of microswimmers, we saw that the emerging interactions between active and passive components are also heterogeneous and anisotropic. In turn, they lead to a complex spatial distribution of colloids and to the possibility of spontaneously de-mixing the suspension. 
These properties are well described by a simple advection-diffusion model for the passive species, based only on the first and second moments of their ensuing active displacements.
From the targeted delivery of chemicals, biomarkers or contrast agents at specific locations in the body to the remediation of water and soils, artificial and biological microscopic self-propelled particles hold a great potential to address critical issues in areas ranging from personalised medical care to environmental sustainability \cite{Wang2012,gao2014}. Development in these areas will depend on our ability to use and control micro/nano-swimmers to perform essential basic functions, such as sensing, collecting and delivering passive cargoes in an autonomous, targeted and selective way \cite{Bechinger2016a}. 
Within this perspective, we have shown here that heterogeneous activity fields can be employed to control the fate of colloidal objects in a manner similar to biological systems, where the spatial modulation of the activity can control, for example, the positioning of organelles within cells \cite{Almonacid2015,Cadot2015,Lin2016}. 
Although this was achieved here through simple confinement, many microswimmers possess -or can be designed to have- the ability to respond to a range  of external physico-chemical cues or stimuli \cite{Kirkegaard2016b,Arrieta2017,Frangipane2018,Arlt2018}. Our work paves the way to using these to dynamically manipulate the fate of colloidal cargoes by externally altering microswimmers' dynamics. 
Advance in this area will require to understand how to predict the active displacements of passive cargoes directly from the dynamics of microswimmers \cite{Cammann2020}, a theoretical development that we leave for a future study.

\bibliography{williams21Biblio}
\bibliographystyle{myh-physrev3}

\begin{acknowledgments}
RJ would like to thank Denis Bartolo, Jean-Fran\c{c}ois Rupprecht and J\'er\'emie Palacci for enlightening discussions. This work was supported in part through the Ram\'on y Cajal Program (RYC-2018-02534; MP), the Spanish Ministry of Science and Innovation (PID2019-104232GB-I00; MP and IT), the Margalida Comas Program (PD/007/2016; RJ), the Juan de la Cierva Incorporaci\'on Program (IJCI-2017-32952; RJ), the Junior Research Chair Program (ANR-10-LABX-0010/ANR-10-IDEX-0001-02 PSL; RJ).
\end{acknowledgments}

\section{Author contributions}
RJ, IT and MP conceived and designed the experiments. RJ and SW performed the
experiments. RJ, SW and MP analyzed the data. RJ developed the analytical model and SW performed the numerical simulations. All authors discussed the results and wrote the article.

\section{Methods}

\subsection{Microorganism culturing}

Cultures of CR strain CC-125 were grown axenically in a Tris-Acetate-Phosphate medium \cite{Harris2009} at 21~\degree{C} under periodic fluorescent illumination (16h light/8h dark cycles, $100 \mu\text{E} \text{m}^{-2} \text{s}^{-1}$, OSRAM Fluora). Cells were harvested at $\sim5 \times 10^6$cells/ml in the exponentially growing phase, then centrifuged at 1000 r.p.m. for 10 min and the supernatants replaced by DI-water containing the desired concentration of polystyrene colloids (Polybead\textsuperscript \textregistered ~Carboxylate Microspheres, radius $a=(5\pm0.5)\,\mu$m, Polysciences).

\subsection{Microfluidics and Microscopy - fill and check details}
The mixed colloid-swimmer solution was injected into polydimethylsiloxane-based microfluidic channels, previously passivated with a $0.5$\%~w/w Pluronic-127 solution to prevent sticking of the particles to the surfaces of the chips. The microfluidic devices were then sealed with Vaseline to prevent evaporation. The channels of varying widths were observed under bright-field illumination on a Zeiss AxioVert 200M inverted microscope. A long-pass filter (cutoff wavelength 765nm) was added to the optical path to prevent phototactic responses of the cells. For each channel width stacks of 80000 images at $10 \times$ magnification were acquired at 10fps with an IDS UI-3370CP-NIR-GL camera.

\subsection{Data analysis}
Particle Trajectories were digitised using a standard Matlab particle tracking algorithm (The code can be downloaded at {\tt http://people.umass.edu/kilfoil/downloads.html}). After the particle trajectories were reconstructed, the large jumps were isolated using a combination of move size and directional correlation along the trajectory (see detailed protocol in \cite{Jeanneret2016, Mosby2020}). 

\subsection{Numerical Simulations}
\label{methods:simu}
One hundred trajectories composed of 10,000 time-steps (with $\delta{t}=0.1\,$s, equal to the acquisition period of the experiments) were simulated for each condition. At each time step an acceptance-rejection scheme is performed to decide whether the particle undergoes a jump: i) a (uniform) random number is first drawn from the interval [0,1], ii) if this number sits within the interval $[(1-\lambda(y)\delta t)/2; (1+\lambda(y)\delta t)/2]$, the particle undergoes a jump whose size is taken from the experimental distribution $q_y(y_{\rm J})$ (using inverse transform sampling), iii) otherwise the particle simply undergoes Brownian motion with diffusivity $D_0 = 0.0439\,\mu$m$^2$s$^{-1}$ (the theoretical bulk diffusivity for a $5\,\mu$m-radius particle at room temperature). Upon reaching a boundary jump moves undergo a stopping boundary condition, stopping at $\pm 5\,\mu$m (the particle's radius) from the boundary (this is to be as faithful with our experimental observations as possible). For simplicity, diffusive moves that cross the boundary undergo a reflective boundary condition. More accurate boundary implementations for the diffusive motion do not lead to a different results as the impact of thermal diffusivity in the channel experiments is negligible. 

\newpage

\section*{Supplementary materials}

\renewcommand{\thetable}{S\arabic{table}}

\setcounter{table}{0}
\renewcommand{\arraystretch}{1.5}

\renewcommand{\thesection}{S\arabic{section}}

\setcounter{section}{0}

\section{Jump detection and effective diffusivity in the no-jump part of the dynamics}
In this section we outline how the jumps are extracted from the trajectories of the colloids and how the parts of the trajectories that do not pertain to jumps are then used to estimate the effective diffusivity in the no-jump part of the dynamics.  

In order to recognise the jump events, we use a method which is similar to both \cite{Mosby2020,Jeanneret2016}, but with slight modifications. The process follows first the method of \cite{Mosby2020}, using directional correlation between subsequent particle displacements to identify candidate periods of non-Brownian displacement. Subsequently, as in \cite{Jeanneret2016}, the individual displacements during these periods are evaluated to ensure that the steps are sufficiently large, further confirming them as non-Brownian in origin.

To look for the directionally correlated windows we start by defining the displacements $\Delta\mathbf{r}(t) = \mathbf{r}(t+\Delta{t}) - \mathbf{r}(t)$, where $\mathbf{r}(t)$ is the colloid position at time $t$. 
With these we calculate the scalar product $p_{\Delta{t}}(t) =\Delta\mathbf{r}(t+\Delta{t}) \cdot \Delta\mathbf{r}(t)$.
In the Brownian case $\langle p_{\Delta{t}}(t)\rangle=0$ with a standard deviation given by $q_c = \sqrt{8}D_0\Delta{t}$. For a diffusivity  $D_0=0.0439\,\mu$m$^2/$s, the thermal diffusivity for a $5\,\mu$m radius at $300\,$K in bulk water, and a time between frames $\Delta{t}=0.1\,$s, we have $q_c = 0.0124\,\mu$m$^2$. Notice that, due to hydrodynamic interactions between colloids and boundaries, $D_0$ over-estimates the thermal diffusivity of the colloids in our experiments. This makes the threshold for classification of active jumps more stringent. 
From this we compute $q_{\Delta{t}}(t) = (p_{\Delta{t}}(t+\Delta{t}) + p_{\Delta{t}}(t) )/2$, and use $4q_c$ as a threshold to select sections of a colloids' trajectory which are sufficiently directionally correlated, and could therefore be active jumps.  Finally, we confirm that each of these potential jumps (consecutive time steps where  $q_{\Delta{t}}(t)$ is above the threshold) has individual displacements which are sufficiently large to rule out correlated Brownian motions. To do this, we calculate for these trajectories the displacement magnitudes $|\Delta\mathbf{r}(t)| = |\mathbf{r}(t+1) - \mathbf{r}(t)|$, we can then compare these values to the Brownian expectation $\sqrt{4D_0\Delta{t}} = 0.132\,\mu$m. 
For a candidate jump trajectory, we require that at least $70\%$ of the individual displacements within it are larger than the Brownian prediction for the overall jump to be confirmed. This percentage strikes a balance between being overselective -losing a large portion of actual jumps- and underselective -mis-classifying non-jump events as jumps. We have thoroughly tested this selection procedure by comparing the experimental movies with the proposed classification outlined above. 

Once the jump sections of the trajectories have been recognised, it is then possible to estimate $D_{WJ}$, the diffusivity that the colloids would have in absence of jumps. This returns $D_{WJ}\simeq 0.162\pm0.0057\,\mu$m$^2/$s. As expected, this is larger than the thermal contribution as the active displacements are not all jumps (see \cite{Leptos2009c,Jeanneret2016}).

\section{Poissonian approximation of the waiting time distributions}
Figure~2a in the main manuscript shows a semilog plot of the waiting time distributions for the $W=100\,\mu$m case, for 10 uniformly spaced bins within the range of $y$-values that goes from the channel centre to the closest approach of the colloids to the side walls. 
The distributions for the other values of $W$ are equivalent. All distributions share the same features. 
They are well described by a single exponential (straight line in semilog) beyond  $\approx 3\,$s, but extrapolating this exponential to shorter times underestimates the actual number of events. 
As can be seen in the Supplementary Movie~M3, the fundamental reason for this deviation is that, in some instances, after the initial interaction between an alga and a colloid, the colloid can alter the motility of the cell in a way that leads it to turn and interact again with the colloid. 
This generates an excess of interactions at short  time, over those that would be estimated by a simple Poissonian process. 
Still, the process is only weakly non-Poissonian and for modelling purposes we will still describe the waiting time dynamics within each bin in terms of a single effective $y$-dependent rate $\lambda(y)$ and a distribution of active jumps that is independent of the waiting time elapsed. However, we need to find a reasonable way to extract these rates from the waiting time distribution. To do that, we start by considering the colloidal trajectories within a $120\,\mu$m-wide strip across the central position $y_{ctr}$ of the $2W=200\,\mu$m-wide channel. 
In this area the effective motion of the colloids is diffusive with a spatially constant diffusivity  which we measure experimentally to be $D_{\text{eff}}=1.82 \pm 0.02\,\mu$m$^2/$s. 
We assume that, besides thermal diffusion, the colloids' motion can be described in terms of active isotropic jumps of magnitude sampled independently from $P(\ell,y_{ctr})$  (see Fig.~S4a) with a Poissonian waiting time dynamics with rate  $\lambda(y_{ctr})$. 
The effective long-time diffusivity is then given by $D_{\text{eff}}=D_{WJ}+\frac{1}{2}\lambda(y_{ctr})\langle\ell^2(y_{ctr})\rangle$ . 
Considering $D_{WJ}\simeq 0.162\,\mu$m$^2/$s and the experimental distribution of jump magnitudes, giving $\langle\ell^2(y_{ctr})\rangle\simeq 15.5\,\mu$m$^2$, the measured value of $D_{\text{eff}}$ can be used to estimate the value of $\lambda(y_{ctr})$. We obtain $\lambda(y_{ctr})\simeq 0.23\,$s$^{-1}$. 
This value should be compared with $\lambda(y_{ctr})_{lin}=0.97\pm 0.02\,$s$^{-1}$ obtained from the standard least squares fitting of the waiting time distribution, and with $\lambda(y_{ctr})_{log}=0.32\pm 0.01\,$s$^{-1}$ which instead is obtained by least squares fitting of the logarithm of the waiting time distribution. We see that the latter, which gives more weight to the part of the waiting time distribution beyond the peak, provides a much better estimate for the interaction rate.  We have therefore decided to extract the effective rates $\lambda(y)$ by least squares fitting of the logarithm of the experimental waiting time distributions (Fig.~2a, solid lines). 

\section{Experimental concentrations of {\it Chlamydomonas} cells across the experiments}

Although we strived for a constant concentration of cells across all experiments, this cannot be controlled exactly. As a result, there were small differences in the measured 2D  cell number concentration between different experiments. This has an effect on the rate of active encounters which, at these concentrations, is expected to be proportional to cell concentration \cite{Jeanneret2016}. The cell concentrations are given in Table~\ref{Table:cell_conc}.

\begin{table}[h]
\addvbuffer[12pt]
{\begin{tabular}{| c | c | c |}
    \hline
    Channel type & Channel Size $\mu$m & Cell Surface Density (cells/$\mu$m$^2$) ($\times 10^{-4}$) \\ \hline
    Straight & 50 & 2.7 $\pm$ 0.7 \\ \hline
    Straight & 100 & 4.5 $\pm$ 0.8 \\ \hline
    Straight & 150 & 2.9 $\pm$ 0.6 \\ \hline
    Straight & 200 & 4.0 $\pm$ 0.7 \\ \hline
    Circular & 200 & 3.8 $\pm$ 1.4 \\ \hline
  \end{tabular}}
\caption{Experimental cell density for the different experiments.}
\label{Table:cell_conc}
\end{table}

\section{Fitting of the data used for the analytical model}

The experimental curves $\Lp(y)$, $\Lm(y)$ and $\lambda(y)$ have been fitted by the following analytical functions: 
\begin{align}
\Lp(y)&=(\Lpw-\Lb)e^{-\frac{y-y_{\rm col}}{l^+}}+\Lb, \label{fit:Lp}\\
\Lm(y)&=(\Lmw-\Lb)e^{-\frac{y-y_{\rm col}}{l^-}}+\Lb, \label{fit:Lm}\\
\lambda(y)&=(\lambda^w-\lambda^b)e^{-\frac{y-y_{\rm col}}{l^{\lambda}}}+\lambda^b,\label{fit:lambda} 
\end{align}
with $\Lpw=\Lmw=0.375\,\mu$m fixed. The best fits (dotted lines in Fig. 2 and 3) are given by the parameters in Table~\ref{Table_Lfit}. 
\begin{table}[h]
\addvbuffer[12pt]{\begin{tabular}{| c | c |}
    \hline
    $\Lb$ & $3.6$  \\ \hline
    $l^+$ & $5.7$  \\ \hline
    $l^-$ & $10.9$ \\ \hline
    $\lambda^w$ & $0.19$  \\ \hline
    $\lambda^b$ & $0.32$  \\ \hline
    $l^{\lambda}$ & $8.33$  \\ \hline
  \end{tabular}}
\caption{Fit parameters for Eqs.~\ref{fit:Lp},\ref{fit:Lm},\ref{fit:lambda}}
\label{Table_Lfit}
\end{table}
The experimental curves $\Lp(y)$ and $\Lm(y)$ were obtained from the experimental jump size distributions $q_{y}(y_{\rm J})$ by fitting the logarithm of each side (i.e. positive and negative $y_{\rm J}$) with affine functions. Because these distributions deviate from simple exponentials as we get closer to the boundary (Fig.~2d), the fits were restricted to $|y_{\rm J}|<{\rm lim}(y)$ in order to extract the characteristic length scales of the initial decay of the distributions. The resulting fits are shown in Fig.~2d (solid lines). The limits for the fits of the experimental distributions were fixed as in Table~\ref{Table_lambda_limits}. 
\begin{table}
\addvbuffer[12pt]{\begin{tabular}{| c | c | c | c | c | c | c | c | c | c | c | c | c | c | c |}
    \hline
    $y~({\rm \mu m})$ (bin center) & $5.0$ & $8.3$ & $11.7$ & $15.0$ & $18.4$ & $21.7$ & $25.0$ & $28.4$ & $31.7$ & $35.0$ & $38.4$ & $41.7$ & $45.0$ & $48.4$\\ \hline
    ${\rm lim}(y)~({\rm \mu m})$ & $7$ & $12$ & $12$ & $15$ & $15$ & $18$ & $20$ & $20$ & $20$ & $20$ & $20$ & $20$ & $20$ & $20$ \\ \hline
\end{tabular}}
\caption{Limits in $y_J$ for the $y$-dependent distributions of jumps perpendicular to the channel boundaries.}
\label{Table_lambda_limits}
\end{table}

\section{Derivation of the Fokker-Planck equation (Eq.~2)}
Motivated by the numerical validation of the jump-diffusion dynamics, we aim to develop a general analytical model that accounts for the experimental results. Following Denisov and Bystrik \cite{Denisov2019a} we first write a stochastic (Langevin) equation for the colloidal dynamics: 
\begin{align}
Y_{t+\tau}=Y_t + N_{\tau} + \Delta_{\tau}(Y_t)
\end{align}
where $\tau$ is an infinitesimal time interval, $\Delta_{\tau}$ is a random variable describing the inhomogeneous Poisson process and $N_{\tau}$ is a standard Wiener process. This can then be used to derive the Fokker-Planck equation (Eq.~2). We first write
\begin{align}
\begin{split}
\label{eq:def_prob}
P_{t+\tau}(y) &=\langle \delta (y-Y_{t+\tau})\rangle, \\ 
&=\langle \delta (y-(Y_t + N_{\tau} + \Delta_{\tau}(Y_t))\rangle, \\ 
\end{split} 
\end{align}
with the angular brackets denoting an average over all noise realizations, which we can express in the following way
\begin{align}
\begin{split}
\label{eq:transition}
P_{t+\tau}(y)=\int_{-\infty}^{+\infty} dy' P_t(y') \int_{-\infty}^{+\infty}\int_{-\infty}^{+\infty} dy_{\rm J}\,d\Delta\eta \;p_{\tau}^{y'}(y_{\rm J}) r_{\tau}(\Delta\eta) \delta(y-(y'+y_{\rm J}+\Delta\eta)),
\end{split} 
\end{align}
where $p_{\tau}^{y'}(y_{\rm J})$ is the probability density that $\Delta_{\tau}=y_{\rm J}$ at position $y'$ and $r_{\tau}(\Delta\eta)=e^{-\Delta\eta^2/(4D\tau)}/\sqrt{4\pi D\tau}$ is the transition probability for the Wiener process (with diffusion coefficient $D$) \cite{Denisov2009}. Integrating first over $\Delta \eta$ we get: 
\begin{align}
\begin{split}
\label{eq:transition_2}
P_{t+\tau}(y)=\int_{-\infty}^{+\infty} dy' \frac{P_t(y')}{\sqrt{4\pi D\tau}} \int_{-\infty}^{+\infty}dy_{\rm J}\,p_{\tau}^{y'}(y_{\rm J}) e^{-\frac{(y-y'-y_{\rm J})^2}{4D\tau}},
\end{split} 
\end{align}
which we can Fourier-transform
\begin{align}
\begin{split}
\label{eq:Fourier}
\tilde{P}_{k}(t+\tau)&=\int_{-\infty}^{+\infty} dy' \frac{P_t(y')}{\sqrt{4\pi D\tau}} \int_{-\infty}^{+\infty}dy_{\rm J}\,p_{\tau}^{y'}(y_{\rm J})\int_{-\infty}^{+\infty} dy\, e^{-iky}e^{-\frac{(y-y'-y_{\rm J})^2}{4D\tau}}, \\
&=\int_{-\infty}^{+\infty} dy' P_t(y')\int_{-\infty}^{+\infty}dy_{\rm J}\,p_{\tau}^{y'}(y_{\rm J}) e^{-ik(y'+y_{\rm J})}e^{-D\tau k^2}.
\end{split} 
\end{align}
Now we are going to expand Eq.~\ref{eq:Fourier} at first order in $\tau$. Following \cite{Denisov2019a}, we can write 
\begin{align}
\begin{split}
p_{\tau}^{y'}(y_{\rm J})= (1-\lambda(y')\tau)\delta(y_{\rm J})+\lambda(y')\tau q_{y'}(y_{\rm J})+o(\tau^2),
\end{split}
\end{align}
where $q_{y'}(y_{\rm J})$ is the probability of having a jump of size $y_{\rm J}$ at position $y'$ and $\lambda(y')$ is the Poissonian rate at position $y'$. We can then get the first order expansion: 
\begin{align}
\begin{split}
\tilde{P}_{k}(t+\tau)&\approx(1-Dk^2\tau)\int_{-\infty}^{+\infty} dy'\,P_t(y') e^{-iky'}-\tau\int_{-\infty}^{+\infty} dy'\,P_t(y') \lambda(y') e^{-iky'}\\
&+\tau\int_{-\infty}^{+\infty}\int_{-\infty}^{+\infty} dy_{\rm J}\,dy'\, P_t(y') \lambda(y') q_{y'}(y_{\rm J})e^{-ik(y'+y_{\rm J})}, \\
&\approx(1-Dk^2\tau)\tilde{P}_{k}(t)-\tau\int_{-\infty}^{+\infty} dy' P_t(y') \lambda(y') e^{-iky'}\\
&+\tau\int_{-\infty}^{+\infty}\int_{-\infty}^{+\infty} dy_{\rm J}dy' P_t(y') \lambda(y') q_{y'}(y_{\rm J})e^{-ik(y'+y_{\rm J})}.
\end{split}
\end{align}
From this expansion, we can express the time-derivative of $\tilde{P}_{k}(t)$
\begin{align}
\begin{split}
\label{eq:last_step}
\frac{\partial \tilde{P}_{k}(t)}{\partial t}&=\lim_{\tau \to 0} \frac{\tilde{P}_{k}(t+\tau)-\tilde{P}_{k}(t)}{\tau},\\
&=-Dk^2\tilde{P}_{k}(t)-\int_{-\infty}^{+\infty} dy' P_t(y') \lambda(y') e^{-iky'}\\
&+\int_{-\infty}^{+\infty}\int_{-\infty}^{+\infty} dy_{\rm J}\,dy'\, P_t(y') \lambda(y') q_{y'}(y_{\rm J})e^{-ik(y'+y_{\rm J})}. \\
\end{split}
\end{align}
We can now inverse Fourier-transform Eq.~\ref{eq:last_step} to get
\begin{align}
\begin{split}
\frac{\partial P_{t}(y)}{\partial t}&=D\frac{\partial^2 P_{t}(y)}{\partial y^2}-\int_{-\infty}^{+\infty} dy' P_t(y') \lambda(y') \delta(y-y')\\
&+\int_{-\infty}^{+\infty}\int_{-\infty}^{+\infty} dy_{\rm J}dy' P_t(y') \lambda(y') q_{y'}(y_{\rm J})\delta(y-y'-y_{\rm J}), \\
\end{split}
\end{align}
and finally obtain Eq.~2
\begin{align}
\label{eq:Fokker-Planck}
\frac{\partial P_{t}(y)}{\partial t}&=D_0\frac{\partial^2 P_{t}(y)}{\partial y^2}- \lambda(y)P_t(y)+\int_{-\infty}^{+\infty}  \lambda(y-y_{\rm J}) P_t(y-y_{\rm J})q_{y-y_{\rm J}}(y_{\rm J})dy_{\rm J}.
\end{align}

\section{Derivation of the drift-diffusion equation (Eq.~3): Kramers-Moyal expansion}

\noindent From Eq.~\ref{eq:Fokker-Planck} above, we can perform a Kramers-Moyal expansion in order to get an effective drift-diffusion equation (Eq.~3). We have (rewriting $q_{y-y_{\rm J}}(y_{\rm J})=q(y-y_{\rm J},y_{\rm J})$)
\begin{align}
\lambda(y-y_{\rm J})P_t(y-y_{\rm J})q(y-y_{\rm J},y_{\rm J})&=\lambda(y)P_t(y)q(y,y_{\rm J})-y_{\rm J}\frac{\partial}{\partial y}\Big[ \lambda(y)P_t(y)q(y,y_{\rm J})\Big] \\
&+\frac{y_{\rm J}^2}{2}\frac{\partial^2}{\partial y^2}\Big[\lambda(y)P_t(y)q(y,y_{\rm J})\Big]+o(y_{\rm J}^3)
\end{align}
which leads to
\begin{align}
\frac{\partial P_{t}(y)}{\partial t}&=D\frac{\partial^2 P_{t}(y)}{\partial y^2}- \lambda(y)P_t(y)+\lambda(y)P_t(y)\int_{-\infty}^{+\infty}  q_{y}(y_{\rm J})dy_{\rm J} \\
&-\frac{\partial}{\partial y}\Big[ \lambda(y)P_t(y)\int_{-\infty}^{+\infty} y_{\rm J}q_{y}(y_{\rm J})dy_{\rm J}\Big]
+\frac{\partial^2}{\partial y^2}\Big[\frac{\lambda(y)P_t(y)}{2}\int_{-\infty}^{+\infty} y_{\rm J}^2q_{y}(y_{\rm J})dy_{\rm J}\Big],
\end{align}
where we recognize the moments of the distribution of jump size $q_{y}(y_{\rm J})$, $m_n(y)=\int_{-\infty}^{+\infty}y_{\rm J}^n q_{y}(y_{\rm J})dy_{\rm J}$. 
\newline
Since $m_0(y)=1$ for all $y$'s, we have
\begin{align}
\begin{split}
\label{approx_unbounded_FP}
\frac{\partial P_{t}(y)}{\partial t}&=\frac{\partial}{\partial y}\Big[\Big(D+\frac{\lambda(y)m_2(y)}{2}\Big)\frac{\partial P_t}{\partial y}\Big]\\
&-\frac{\partial}{\partial y}\Big[\Big(\lambda(y)m_1(y)-\frac{1}{2}\frac{\partial}{\partial y}[\lambda(y)m_2(y)]\Big)P_t(y)\Big]
\end{split}
\end{align}
which is a drift-diffusion equation with effective diffusivity $D_{\rm eff}(y)=D+\frac{\lambda(y)m_2(y)}{2}$ and effective drift $V_{\rm eff}(y)=\lambda(y)m_1(y)-\frac{1}{2}\frac{\partial}{\partial y}[\lambda(y)m_2(y)]$.

\section{Filling dynamics in the demixing experiments}
In the demixing experiments, let us call $N_c(t)$ and $N_s(t)$ the total number of colloidal particles that at time $t$ are in the circular chamber or in the side channels respectively. In a first-order kinetics, these obey the following linear evolution:
\begin{align}
\frac{d N_c(t)}{dt} &= -k_{\rm in} N_c(t) + k_{\rm out}N_s(t)\\
\frac{d N_s(t)}{dt} &= +k_{\rm in} N_c(t) - k_{\rm out}N_s(t).
\end{align}
Notice that this assumes that the total number of colloids in the chamber and the side channels, $N_t = N_c+N_s$, is constant in time. This is what we observe in our experiments (purple solid line in Fig.~4b). This set of equations is immediately solved to give
\begin{align}
N_c(t) &= N_t\left[  \frac{k_{\rm out}+k_{\rm in}e^{-(k_{\rm out}+k_{\rm in})t}}{k_{\rm out}+k_{\rm in}}\right]\\
N_s(t) &= N_t\left[  \frac{k_{\rm in}\left(1-e^{-(k_{\rm out}+k_{\rm in})t}\right)}{k_{\rm out}+k_{\rm in}}\right].
\end{align}

\section{Estimating the rate of escape from a circular chamber into the side channels: $k_{in}$}

We will outline two methods of differing complexity that can be used to predict the escape rate of the colloids $k_{\text{in}}$, whose experimentally measured value is $k_{\rm in}=(7.8\pm 0.6)\times 10^{-5}\,$s$^{-1}$.
The system we consider is composed of a single particle within a circular chamber, subject to diffusion and -later- drift. 
The walls of the circular chamber divided into two types: the first is a no-flux part which the colloids cannot penetrate; the second is an absorbing part where the colloids are removed from the system. We will estimate the escape rate as the inverse of the average time taken by a colloid to be absorbed at the boundary. This is of course a version of the famous `Narrow Escape Problem' \cite{redner2001} with two variations. Firstly, the particles are subject to a space-dependent diffusivity and drift, while the Narrow Escape approaches generally have constant diffusivity and no drift. Secondly, in order to stay faithful to the geometry of the experiments, the boundary is composed of several distinct absorbing patches, rather than a single one (of the same total size) as would be standard in the Narrow Escape Problem. 

For a single absorbing patch at the boundary of a disk, the Narrow Escape Problem has been solved for a particle with constant diffusivity in \cite{Singer2006}. Following this work we can begin noting that solving the mean first passage time, and hence escape rate, means solving the following Poisson equation with mixed Neumann-Dirichlet inhomogeneous boundary conditions:
\begin{equation}
\begin{cases}
D \Delta t(r,\theta) = -1& \text{for}\; r < R,\,0 \leq \theta < 2\pi\\
t(r,\theta) = 0 & \text{for}\; r = R, \, \theta \in \Theta_a\\
\frac{\partial t(r,\theta)}{\partial r} = 0& \text{for}\, r = R,\, \theta \notin \Theta_a
\end{cases}
\label{eq:lambda_D_const}
\end{equation}
where $(r,\theta)$ are the coordinates on the disk of radius $R$, $D$ the constant diffusivity, $t(r,\theta)$ the escape time given initial position $(r,\theta)$ and $\Theta_a$ is the set of angles for which the boundary is absorbing. In this case this is a set of 12 regions with angles corresponding to $7.5\mu$m exits.
This set of equations can then be solved numerically for a given prescribed boundary and diffusivity to give the escape rates of the colloids.

\begin{table}[ht]
\centering
\addvbuffer[12pt]{\begin{tabular}{| c | c | c |c |}
    \hline
    $D$ ($\mu$m$^2$s$^{-1}$) & $t(0,0)^{-1}$ ($\times 10^{-5}$ s$^{-1}$) & $\langle{t(r,\theta)}\rangle^{-1}$ ($\times 10^{-5}$ s$^{-1}$) & $\langle{t(r,\theta)}\rangle_\rho^{-1}$ ($\times 10^{-5}$ s$^{-1}$) \\ \hline
    $D_{\text{th}}$ =  0.05 & 0.228 & 0.295 & 0.280 \\ \hline
    $D_{\text{bulk}}$ =  3.55 & 16.2 & 20.8 & 19.9 \\ \hline
    $\langle{D(r,\theta)}\rangle$ = 3.14 & 14.3 & 18.3 & 17.6 \\ \hline
    $\langle{D(r,\theta)}\rangle_\rho$ = 3.12 & 14.2 & 18.3 & n/a \\ \hline
  \end{tabular}}
\caption{Escape rates for constant diffusivities and no drift.}
\label{Table_fixed_diff}
\end{table}
Table~\ref{Table_fixed_diff} shows the rates obtained from the numerical solution of Eq.~\eqref{eq:lambda_D_const} for four different values for the constant diffusivity: i) the bulk thermal value (to be used as a baseline); ii) the bulk effective diffusivity $D_{\text{bulk}}$ predicted by the Kramers-Moyal (KM) model; iii) the spatially-averaged KM diffusivity over the whole system, $\langle{D(r,\theta)}\rangle$; iv) the average KM diffusivity weighted by the stationary colloidal distribution $\rho$, $\langle{D(r,\theta)}\rangle_{\rho}$. For each fixed diffusivity, we report the escape rates calculated with i) a fixed starting point at the centre of the chamber ($t(0,0)^{-1}$); ii) a uniformly distributed initial particle position ($\langle{t(r,\theta)}\rangle^{-1}$); iii) an initial particle position distributed according to $\rho$ ($\langle{t(r,\theta)}\rangle_{\rho}^{-1}$). It is clear that all these rates largely overestimate the experimental one. 

Up to this point we have limited ourselves to a constant diffusivity and no particle drift, but we have seen in the KM model that such features are required to recapitulate the colloidal distributions. In order to include them in the estimate of the escape, we  perform a numerical simulation of a colloid subject to the space-dependent effective diffusivity and drift used in the KM model. The dynamics for each component follows:
\begin{eqnarray}
x(t+\delta{t}) = x(t) + \sqrt{2 D(r) \delta{t}}\, \xi_1(t) + v_1(r) \delta{t}\cos(\theta)  + v_2(r) \delta{t} \cos(\theta) ,\nonumber \\
y(t+\delta{t}) = y(t) + \sqrt{2 D(r) \delta{t}}\, \xi_2(t) + v_1(r) \delta{t}\sin(\theta) + v_2(r) \delta{t} \sin(\theta),
\label{eq:Langevin_escape}
\end{eqnarray}
Here, $(x(t),y(t))$ is the position of the colloid at time $t$, $D(r) = D_0 + \frac{1}{2}\lambda(r)m_2(r)$ is the local effective diffusivity at a position $r = |(x,y)|$, $\mathbf{\xi}(\cdot)$ is a Gaussian white noise of variance 1, $v_1(r) = \lambda(r)m_1(r)$ is the drift due to the first moment of the jump distributions, $v_2(r) = \frac{1}{2}D'(r) \left(\xi_i(t)^2 - \alpha \right)$ the drift due to the second moment. The constant $\alpha$ captures the integration scheme \cite{Lancon2001}, with $\alpha=0,1$ corresponding to the Stratonovitch and It\^o respectively. The angle $\theta = \arctan(y(t)/x(t))$ ensures the drifts are oriented towards the center of the chamber. Of course, simulations allow the boundary conditions to match those in the experiment.

In our simulations we use: $\delta{t} = 0.1$s; the same boundary structure used to estimate the values of Table~\ref{Table_fixed_diff}; and the KM effective diffusivity and drift calculated for the $100\,\mu$m channel rescaled by the ratio of the concentrations between that experiment and the circular chamber one. The results can be found in Table~\ref{Table_alpha} for the escape rate given an initial condition of particles at the centre of the chamber, where the error is the stochastic error from the simulations.

\begin{table}[h]
\addvbuffer[12pt]{\begin{tabular}{| c | c | c |c |}
    \hline
    $\alpha$ & $t(0,0)^{-1}$ ($\times 10^{-5}$ s$^{-1}$) & $\pm$ stochastic error from simulations ($\times 10^{-5}$ s$^{-1}$) \\ \hline
    0 (Stratonovitch) & 7.408 & 0.066  \\ \hline
    1 (Ito) & 3.063 & 0.027 \\ \hline
  \end{tabular}}
\caption{Escape rates calculated from Eq.~\ref{eq:Langevin_escape}.}
\label{Table_alpha}
\end{table}

\section{Supplementary Movies}
The Supplementary Material includes the following movies:
\begin{enumerate}
\item Movie~M1: {\it C. reinhardtii} microalgae and $10\,\mu$m-diameter polystyrene colloids within a set of microfluidic straight channels. Recorded at 10 frames per second; length conversion factor $0.55\,\mu$m per pixel. 
\item Movie~M2: Colloid-alga interaction showing a single colloid with its trajectory indicated as it interacts with several microoganisms. Recorded at 10 frames per second; length conversion factor $0.55\,\mu$m per pixel. 
\item Movie~M3: Colloid-alga interaction influencing the subsequent swimming of the microorganism and causing it rapid successive interactions. Recorded at 10 frames per second; length conversion factor $0.55\,\mu$m per pixel.
\item Movie~M4: Dynamics of side channels filling with colloids due to activity of the microalgae. Recorded at 0.1 frames per second; length conversion factor $0.23\,\mu$m per pixel.
\item Movie~M5: Same as Movie~M4 but without the microalgae in the main chamber. Recorded at 0.25 frames per second; length conversion factor $0.55\,\mu$m per pixel.
\end{enumerate}

\section{Supplementary figures}

\renewcommand{\thefigure}{S\arabic{figure}}

\setcounter{figure}{0}

\begin{figure}[h]
\centering
\includegraphics[width=0.7\linewidth]{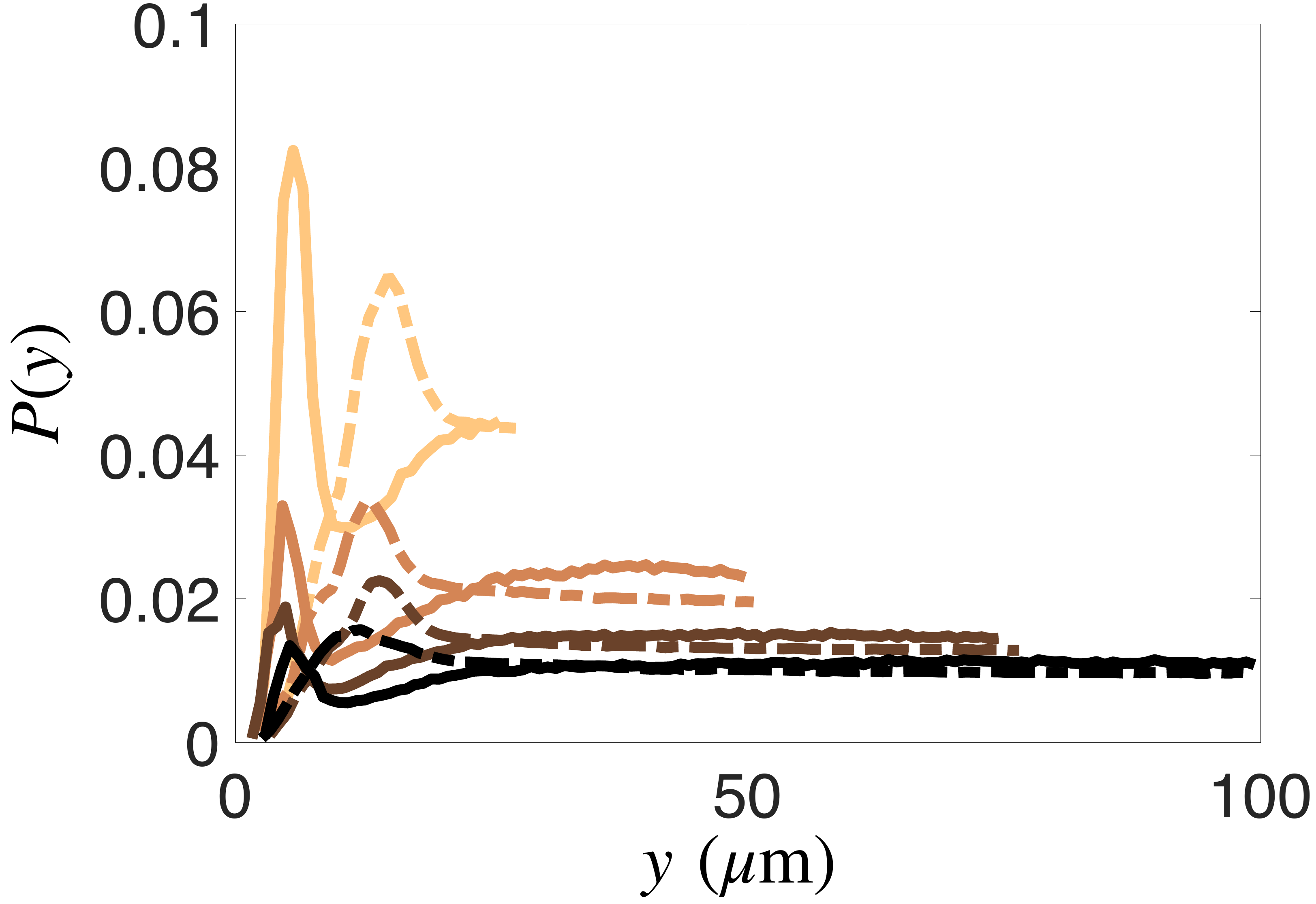}
\caption{Steady state distribution function of algae (dash-dotted lines) and colloids (solid lines, same data as in Fig.~1b) across the straight channels, from the boundary to the channel midpoint. The algal distribution displays a clear peak at $y_{\rm CR}\approx 15~\mu{\rm m}$.}
\label{figS1}
\end{figure}

\begin{figure}[h]
\centering
\includegraphics[width=0.7\linewidth]{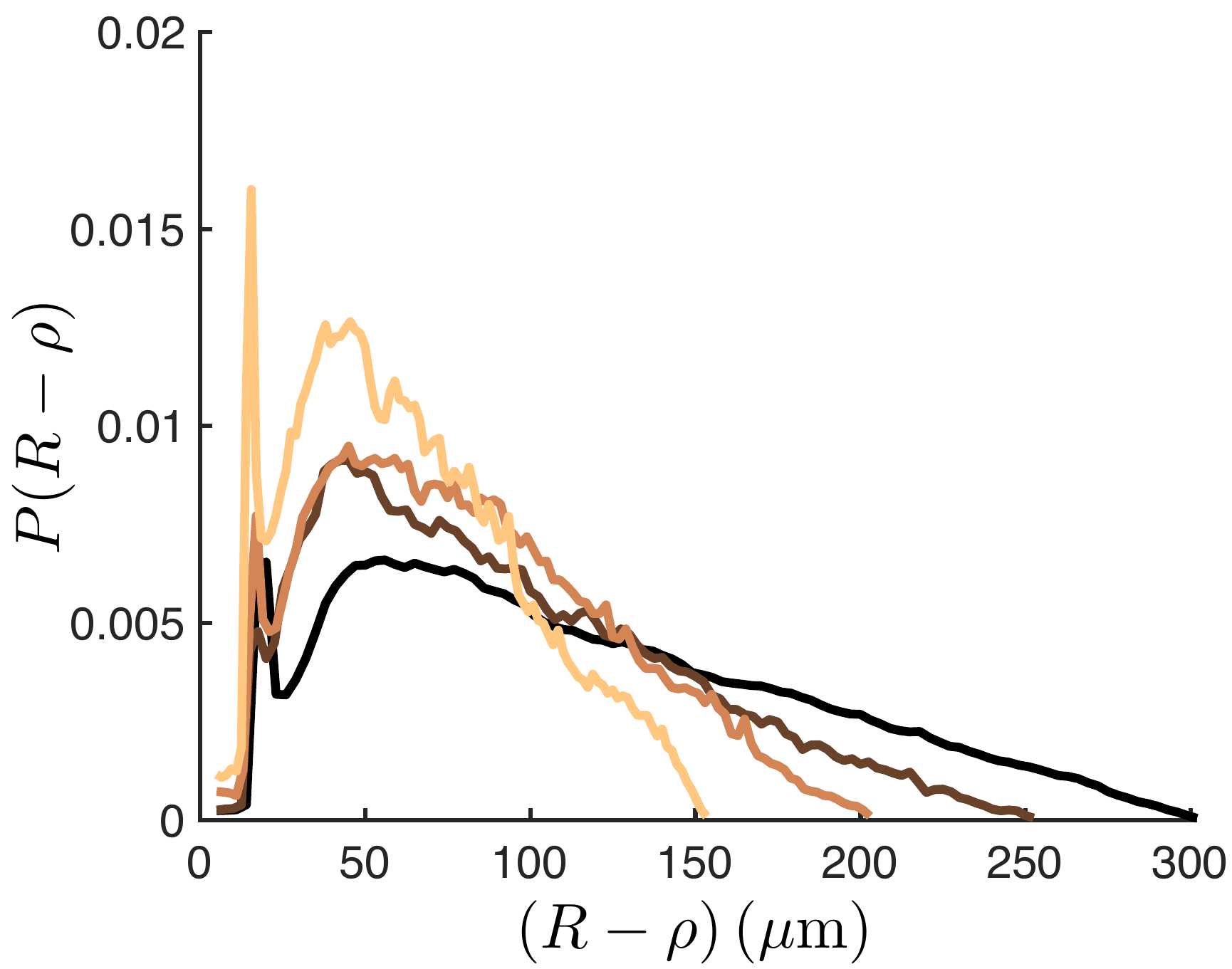}
\caption{Steady state probability distribution function $P(\rho)$ for colloids in active-passive experiments, within circular chambers of radius $R$. By symmetry, the probability distribution function is only a function of the radial coordinate $\rho$. The probability distribution function is defined in such a way that $\int_0^RP(\rho) {\rm d}\rho=1$.}
\label{figS2}
\end{figure}

\begin{figure}[h]
\centering
\includegraphics[width=0.5\linewidth]{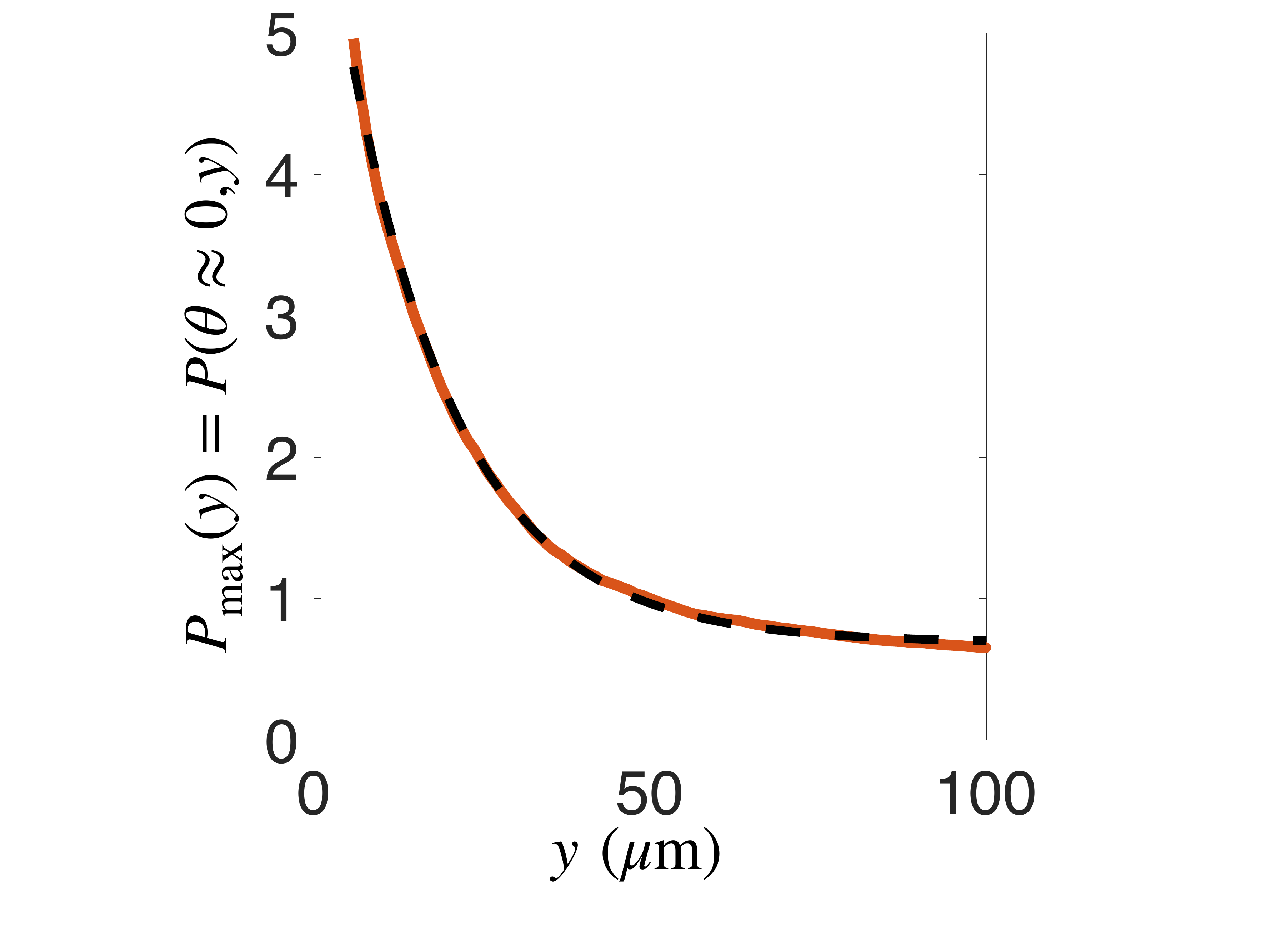}
\caption{Maximum value in the distributions of jump angle $P_{\rm max}(y)=P(\theta\approx 0,y)$ (Fig. 2D) as a function of distance from the wall. The peak value decays exponentially with a characteristic length scale $L_{\theta}=16.4\pm0.3 {\rm \mu m}$.}
\label{figS3}
\end{figure}

\begin{figure}[h]
\centering
\includegraphics[width=\linewidth]{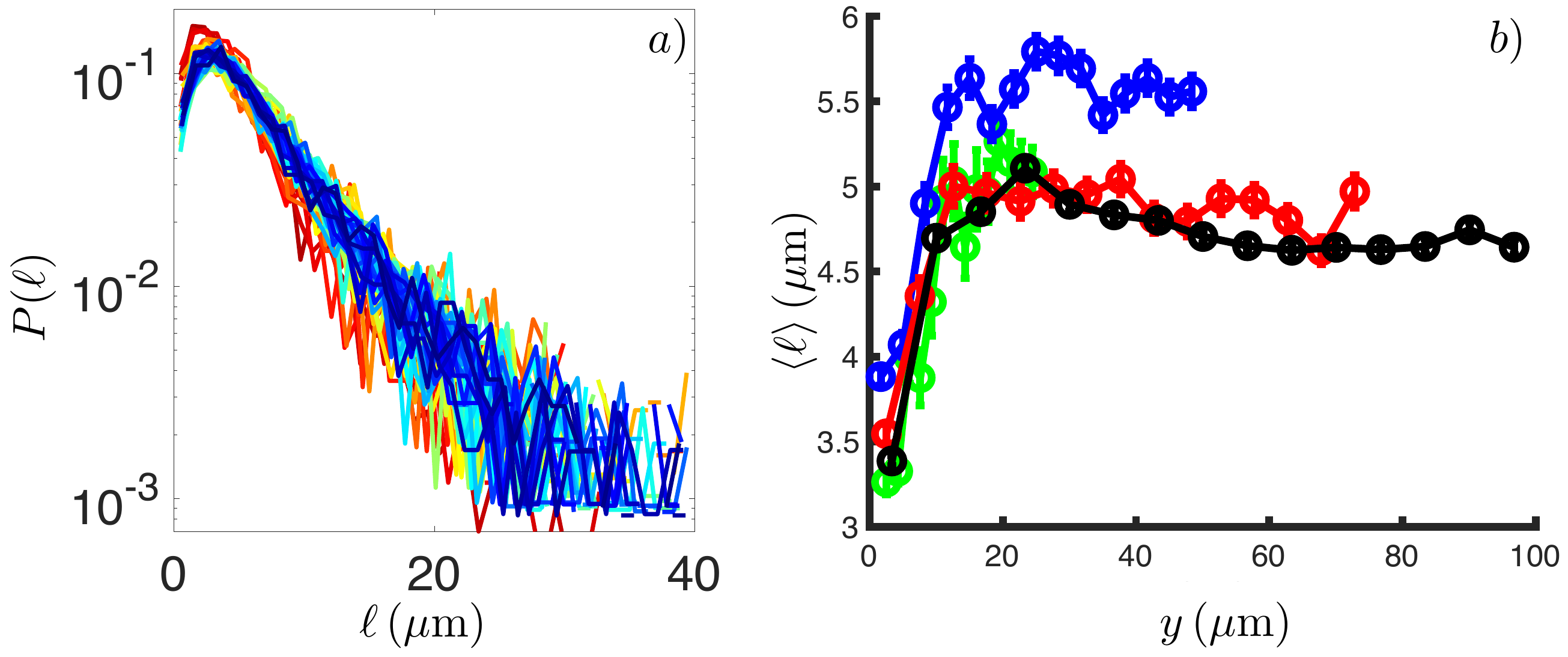}
\caption{a) Probability distribution functions of jump lengths $P(l)$ at different distances $y$ from the channel boundary (in the $2W=100 {\rm \mu m}$ channel). Semi-log plot. The colour code is the same as in Fig.~2 of the main manuscript (red at the boundary, blue in the middle of the channel). b) Position-dependent average jump magnitude $\langle\ell\rangle(y)$ for the four values of $W$ explored ($2W=50,100,150,200\,\mu$m).}
\label{figS4}
\end{figure}

\begin{figure}[h]
\centering
\includegraphics[width=0.5\linewidth]{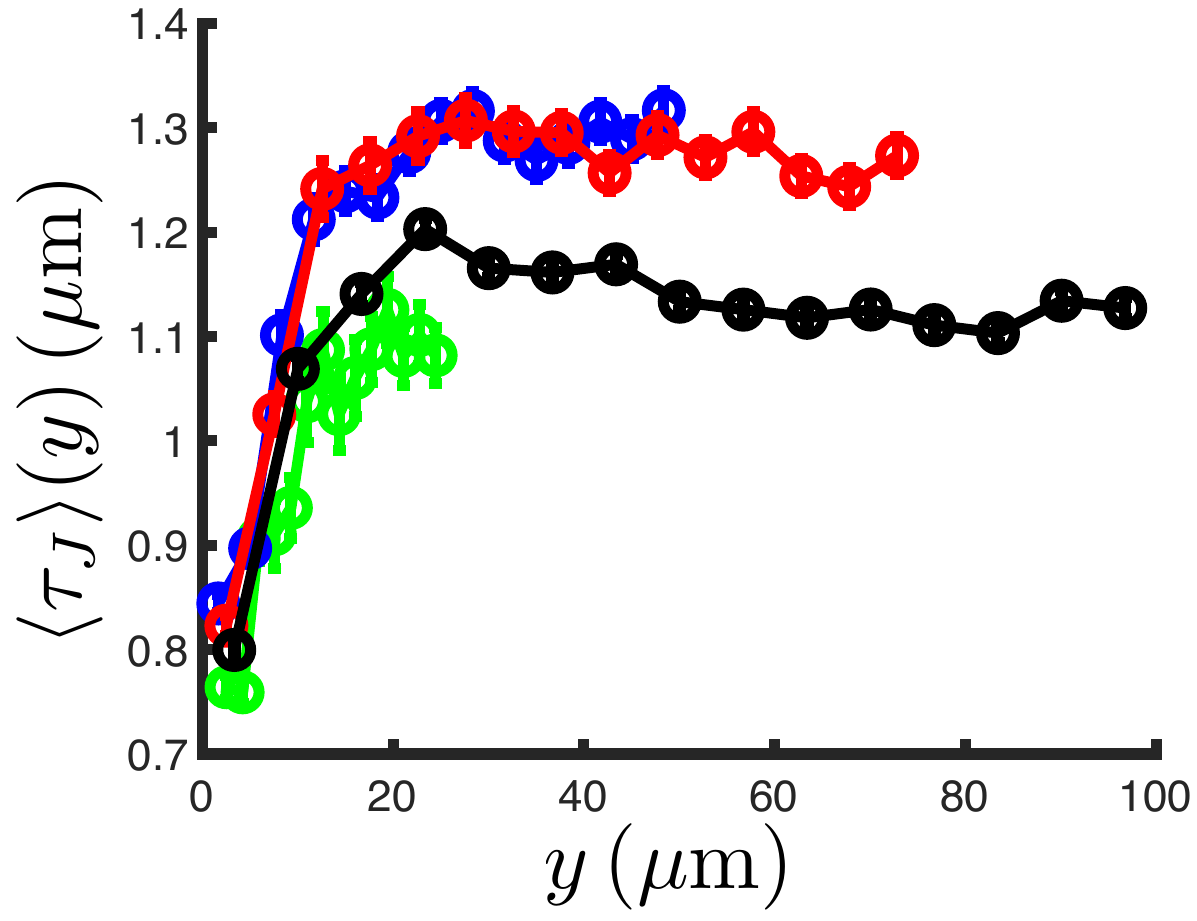}
\caption{Average duration of a jump event $\langle\tau_J\rangle$(y) as a function of distance $y$ from the nearest boundary. The colours (green, blue, red, black) correspond to  the four values of $W$ explored ($2W=50,100,150,200\,\mu$m).}
\label{figS5}
\end{figure}

\clearpage

\end{document}